\documentclass[fleqn,usenatbib,useAMS]{mnras}
\usepackage{graphics,epsf}
\usepackage{amsmath}                
\usepackage{amsfonts}               
\usepackage{amssymb}                
\usepackage{epsfig}                 
\usepackage{graphicx}

\usepackage{xcolor}
\definecolor{redak}{rgb}{0.9,0.15,0.05}

\def \kms{\,\rm{km\,s^{-1}}}

\def \msyr{\,\rm{M_{\odot}}\,\rm{yr^{-1}}}
\def \cm{\,\rm{cm}}

\def \g{\,\rm{g}}

\def \au{\,\rm{au}}
\def \erg{\,\rm{erg}}

\def \yr{\,\rm{yr}}

\def \etc{$\eta$~Car}
\def \days{\,\rm{days}}

\def \rmModot{\,\rm{M_{\sun}}}
\def \rmRodot{\,\rm{R_{\sun}}}

\title[Fast ejecta in $\eta$ Car]{Fast ejecta resulted from jet-wind interaction in the Great Eruption of Eta Carinae}

\author[M. Akashi and A. Kashi]{
Muhammad Akashi$^{1,2,3}$\thanks{Muhammad Akashi e-mail: \href{mailto:akashi@physics.technion.ac.il}{akashi@physics.technion.ac.il}}
and
Amit Kashi$^{1}$\thanks{Amit Kashi e-mail: \href{mailto:kashi@ariel.ac.il}{kashi@ariel.ac.il}}
\\
$^{1}$Department of Physics, Ariel University, Ariel, POB 3, 4070000, Israel \\
$^{2}$Department of Physics, Technion, Haifa 3200003, Israel\\
$^{3}$Kinneret College on the Sea of Galilee, Samakh 15132, Israel\\
}

\date{Accepted XXX. Received YYY; in original form ZZZ}

\pubyear{2020}

\begin{document}
\label{firstpage}
\pagerange{\pageref{firstpage}--\pageref{lastpage}}
\maketitle

\begin{abstract}
The accretion model for the nineteenth century Great Eruption (GE) of $\eta$ Carinae suggests that mass outflowing from the primary was accreted onto the secondary, and the gravitational energy of that mass accounts for the increase in luminosity and most of the kinetic energy of the ejecta.
It further argues that the accretion was accompanied by the ejection of two jets that shaped the bipolar Homunculus nebula.
Observations of echos from the GE found emission lines with broad wings suggesting some of the mass in equatorial directions reached more than $10\,000 \,\rm{km\,s^{-1}}$.
We run hydrodynamic simulations following periastron passage during the GE, launching jets from the secondary as it accreted gas erupted from the primary. We then follow the interaction of the polar jets with the surrounding primary wind, as they accelerate part of the flow to velocities $ > 10\,000 \,\rm{km\,s^{-1}}$ and deflect it towards lower latitudes.
We find that the amount of mass that reached these high velocities during the GE is $M_h \approx 0.02 \,\rm{M_{\odot}}$. This value reaches maximum and then decreases with time.
Our simulations agree with previous results of the accretion model from which we estimate $M_h$ taking into account the energy budget of the GE.
The accretion model can explain the observations of high velocity gas in light echos with the known two stars, and a triple star system is not required.
\end{abstract}


\section{INTRODUCTION}
\label{sec:intro}

The binary system $\eta$ Carinae (\etc) is composed of an evolved very massive star (hereafter the primary), often considered to be a luminous blue variable (LBV) star, and a hotter and less luminous evolved main-sequence (MS) star \citep[hereafter the secondary;][]{Damineli1996, DavidsonHumphreys1997,DavidsonHumphreys2012}.
The system is famous for the two giant eruptions it experienced in the nineteenth century.
The 1838--1856 Great Eruption (GE) was extremely energetic, involving $\approx 10^{50} \erg$, about half of which in radiation and the other half in kinetic energy of the enormous amount of mass that has been ejected \citep{Smithetal2003, HumphreysMartin2012}. This ejecta later formed what today is known as the Homunculus nebula, containing at least $12\rmModot$, and possibly up to $40\rmModot$ \citep{Smithetal2003, SmithFerland2007, KashiSoker2010}.
Another outburst, the Lesser Eruption (LE), occurred in 1887--1895.
It appeared as much less luminous than the GE, as it was enshrouded by the material ejected by its predecessor \citep{HumphreysMartin2012}.
There is also some evidence for another eruption in the first half of the twentieth century, possibly in 1941, that might have created an asymmetric structure, the ``Baby Homunculus'' \citep{Abrahametal2014} that is still buried in the Homunculus and only starts showing itself in recent observations \citep{BordiuRizzo2019}.
Moreover, at least two eruptions earlier than the GE have been identified from proper motion of numerous N-rich knots in the outer ejecta of \etc \citep{Kiminkietal2016}.

Presently the binary system has an highly eccentric orbit \citep[e.g.,][]{Daminelietal1997, Smithetal2004, Davidsonetal2017}, and strong winds \citep{PittardCorcoran2002, AkashiSoker2006, KashiSoker2009a, Kashi2017, Kashi2019}, giving rise to a strong interaction every $5.54$ years during periastron passage known as the spectroscopic event, where many lines experience rapid variability (e.g., \citealt{Corcoranetal2015, Mehneretal2015, Hamaguchietal2016, Grantetal2020}, and references therein).

The physical mechanism behind the two eruptions is still not fully understood, and a number of scenarios have been proposed involving different mechanisms \citep[e.g.,][]{Davidson2012,Davidson2020}.
Among these we can count a scaled-down Type IIn supernova (SN) with ejecta that interacts with the circumstellar gas \citep{Smith2013};
pulsational instabilities that involve deep regions in the code \citep{GuzikLovekin2014};
deposition of energy underneath the stellar surface \citep{Quataertetal2016};
extreme S-Dor variability \citep{Vink2012};
sudden energy injection into the interior \citep{Kashietal2016,Owockietal2019};
a merger event in a triple system with the surviving companion initiating the merger of the inner binary with the Kozai-Lidov effect \citep{PortegiesZwartvandenHeuvel2016};
exchange of partners in a hierarchical triple system via the triple evolution dynamical instability, leading to a merger and kicking out the original primary as a stripped-envelope Wolf-Rayet star on a wide eccentric orbit \citep{LivioPringle1998,Smithetal2018a};
pulsational pair-instability \citep{Woosley2017};
and He-opacity related instability \citep{Jiang2018}.
Most of these explanations, while reasonably account for the results of the eruption, do not provide a full explanation for its \textit{cause}.

Most of the mass ejected in the GE moves in velocities in the range of $400$--$900\kms$ (e.g., \citealt{Weis2012,Kiminkietal2016} and references therein).
Until recently, the highest velocity material in \etc was observed in filaments in the outer ejecta, reaching projected velocity of $5\,000 \kms$.
Some material was also seen reaching $6\,000 \kms$ \citep{Smith2008}.

Evidence for an even faster outflow was revealed, when \cite{Smithetal2018a} used observations of echos from the GE, and found that the H$\alpha$ line shows broad emission line with wings reaching velocities in the range $-10\,000 \kms$ on the blue side to $+20\,000 \kms$ on the red side.
The echo from which these high velocities were inferred, designated EC2, is closer to \etc than the echo EC1 earlier discovered \citep{Restetal2012}.
\cite{Smithetal2018a} therefore associated the observation of this particular echo with the 1845--1858 plateau observed in the historic light curve.
\cite{Smithetal2018a,Smithetal2018b} analyzed echos from different stages of the GE, recovered velocities, and concluded that the GE was powered by shock interacting with circumstellar gas.
Their proposed scenario has two stages: a slow outflow in the decades before the GE, driven by binary interaction that produced a dense equatorial outflow, and later an explosive energy injection that drove interaction with the circumstellar gas, powering the plateau and sweeping slower circumstellar gas into a fast shell that became the Homunculus. 
Consequently, to make the scenario feasible, they suggested a model for the GE of \etc that involves three stars with alternating roles in the system.

\cite{Smithetal2018a} did not give a quantitative estimate of the mass in the fast ejecta. Calculating the amount of mass would require a number of assumptions about the geometry, density, and ionization fraction, and a radiative transfer model. Concluding the amount of mass in the fast wind based on the observed H$\alpha$ flux is therefore difficult and very uncertain.
However, as the total flux in the very broad wings is almost equal to that narrower line core, it is possible to estimate the fraction of mass at high velocities as more than a few percents of the total mass. As a consequence, this estimate also raises the energy budget of the GE.

A different scenario for the GE, that this paper focuses on, is the \textit{binary accretion model}.
Developed by \cite{Soker2001}, the model suggests that the secondary accreted a large fraction of the mass
that was expelled from the primary star in the GE and LE.
\cite{Soker2004,Soker2005,Soker2007}
further conjectured that the GE was caused by disturbances in the
outer boundary of the inner convective region, which expelled
the outer radiative zone, causing the star to lose $\approx 20 \rmModot$ (part of which accreted by the secondary).
A fraction of the mass accreted by the secondary was channeled to blowing bipolar jets.
The accretion rate was high, and the potential lobe of the primary was probably full of gas for the majority of the binary orbit.
The accreted mass released gravitational energy that might account for
the entire extra energy of the GE, while the jets shaped
the bipolar nebula around the star (the `Homunculus')
and supplied part of its kinetic energy.

\cite{KashiSoker2010} showed that when integrating the binary orbit back in time allowing for mass loss from the system and mass transfer from the primary to the secondary, it is possible to match the peaks in the historical light curve during the GE and the LE with periastron passages.
This gave support to the feasibility of the accretion model in accounting for the giant eruptions, and gave support to the idea that the stars are more massive than previously thought.

The occurrence of accretion in the present-day \etc close to periastron passages was first shown in simulations by \cite{Akashietal2013}.
Later simulations with higher resolution and more detailed physics confirmed the conclusion and were able to quantify the amount of accreted mass \citep{Kashi2017,Kashi2019}.
Since accretion takes place even for the much lower primary mass loss rate the system experiences today, it surely took place during the GE, when the primary mass loss rate was $\approx 1000$ times higher.

In this work we simulate jet interaction with the ambient outflowing gas during the GE, and show that the high velocity gas is obtained, with agreement with the resent observations.
In section \ref{sec:simulation} we describe in detail the numerical simulation.
The results, showing the high velocity gas, are presented in section \ref{sec:results}.
Our a summary and discussion are given in section \ref{sec:summary}.

\section{Modeling the fast outflow}
 \label{sec:model}

\subsection{Theoretical estimate}

Recalling the discussion in section \ref{sec:intro} about the uncertainty in the amount of mass in high velocities, we need to explore the consequences of having a large amount of mass in terms of the GE energy and mass budget.
The energy in the GE is $\simeq 10^{50} \erg$ (section \ref{sec:intro}).
The kinetic energy of the high velocity gas somewhat increases the energy budget of the GE.
The kinetic energy needed to power the fast moving gas with mass $M_h$ at velocity $v_h$ is:
\begin{equation}
\Delta E_k = \frac{1}{2} M_h v_h^2 .
\label{eq:DeltaEk}
\end{equation}
According to the accretion model, kinetic energy of outflowing gas comes from gravitational energy of mass $\Delta M_{\rm{acc}}$ accreted onto the secondary,
\begin{equation}
\Delta E_{\rm{acc}} = \frac{GM_2 \Delta M_{\rm{acc}}}{R_2} .
\label{eq:DeltaEacc}
\end{equation}
Taking, following the virial theorem, a fraction $\xi=0.5$ of that energy to convert to the kinetic energy $E_k= \xi E_{\rm{acc}}$ we obtain
\begin{equation}
\Delta M_{\rm{acc}} = \frac{M_h v_h^2 R_2}{2 \xi G M_2} .
\label{eq:DeltaMacc}
\end{equation}

\cite{KashiSoker2010} found that the ratio between the total mass ejected in the GE, and the total mass accreted onto the companion is in the range $\chi \equiv M_{\rm{acc}} /  M_{\rm{ej}} \simeq 0.15$--$0.2$.
We will assume that similarly this ratio applies here, so $\chi \equiv \Delta M_{\rm{acc}} / \Delta M_{\rm{ej}}$.
As mentioned above, the Homunculus mass is $M_{\rm{ej}} = 20$--$40 \rmModot$, thus from this estimate and equation \ref{eq:DeltaMacc}, we derive our expectation to the mass in the fast outflow 
\begin{equation}
\begin{split}
M_h& \approx \frac{2 \chi \xi G M_2 \Delta M_{\rm{ej}}}{v_h^2 R_2} \\
&\simeq 0.005 
\left( \frac{\chi}{0.16} \right)
\left( \frac{\xi}{0.5} \right)
\left( \frac{\Delta M_{\rm{ej}}}{4 \rmModot} \right)
\left( \frac{M_2}{80 \rmModot} \right)\\
&\times \left(\frac{R_2}{20 \rmRodot}\right)^{-1}
\left( \frac{v_h}{10^4 \kms} \right)^{-2} \rmModot.\\
\end{split}
\label{eq:Mh2}
\end{equation}

The calculation above and the estimate in equation \ref{eq:Mh2} is time dependent and the value will change with time.
The change can be in both directions. Lower density gas can be accelerated to large velocities, increasing $M_h$, and at later times this material will eventually collide with slower gas of larger mass, and will result in the Homunculus nebula as we know today, with much lower velocities.

\cite{KashiSoker2010} also obtained, according to the accretion model, that the difference between the mass transferred from the primary to the secondary, to the accreted mass is $\approx 2 \rmModot$ for the entire GE, or $\approx 0.5 \rmModot$ for each periastron passage.
Some of that ``available for disposal'' mass may be ejected in the form of jets (the rest can outflow from the disk or radiate from the star).
Nest, we will examine how the jets interact with the wind of the primary star and accelerate some of it to high velocities.

\subsection{The numerical simulations}
\label{sec:simulation}

\begin{table*}
\centering
\caption{A list of the simulations presented in the paper.
Run naming code:
C$=$ Conventional mass model. M$=$ High mass model.
Columns, from left to right, indicate: run number,  stellar masses, semi-major axis, eccentricity, primary wind density profile, primary wind velocity, mass loss rate in the jets, velocity of the jets, semi-opening angle of the jets, duration of launching the jets, and the resulted amount of high velocity gas at the time it reaches maximum. 
}
\begin{tabular}{lcccccccccc} 
\hline
Run   &$M_1,M_2$      &$\left< a \right>$   &$e$   &$\rho_w$  &$v_w$              &$\dot{M}_{\rm jet}$ &$v_{\rm jet}$       &$\theta_{\rm jet}$    &Duration    &$M(v > 10^4 \kms)$     \\l
      &($\rmModot$)   &($\au$)              &      &          &($\rm{km~s^{-1}}$) &($\msyr$)           &($\rm{km~s^{-1}}$)  &                      &of jets     &($\rmModot$)          \\
\hline

C1 (fiducial)    & 120,30        &16.64     & 0.9  &eq. \ref{eq:rhow1}   &$500$    &0.6      &$3\,000$      & $15^\circ$                & 20 days    & $5.7 \times 10^{-3}$   \\

C2    & 120,30        &16.64     & 0.9  &eq. \ref{eq:rhow1}   &$500$    &0.6      &$3\,000$      & $70^\circ$                & 20 days    & $2.1 \times 10^{-3}$   \\

C3    & 120,30        &16.64     & 0.9  &eq. \ref{eq:rhow1}   &$500$    &0.12     &$3\,000$     & $15^\circ$                 & 20 days    & $1.2 \times 10^{-3}$   \\ 

C4    & 120,30        &16.64     & 0.9  &eq. \ref{eq:rhow1}   &$500$    &1.2      &$3\,000$      & $15^\circ$                & 20 days    & $1.6 \times 10^{-2}$   \\

M1    & 180,70        &19.75    & 0.9   &eq. \ref{eq:rhow1}   &$500$    &0.6      &$3\,000$      & $15^\circ$                & 20 days    & $5.9 \times 10^{-3}$   \\

\hline
\end{tabular}
\label{table:parameters}
\end{table*}


We use the hydrodynamic code \textsc{flash}, described originally in \cite{Fryxell2000}, but vastly updated since then.
Our 3D Cartesian grid extends over $(L_x,L_y,L_z) = \pm 50 \au$.
The initial conditions are set $2$ days before periastron.
We place the secondary on an eccentric $e=0.9$ Keplerian orbit.

The masses of the two stars are not well constrained, and there are mainly two suggestion for the present masses:
\begin{enumerate}
\item
\emph{Conventional mass model}, where the primary and secondary masses are $M_1=120 \rmModot$ and $M_2=30 \rmModot$, respectively \citep{Hillieretal2001}.
\item \emph{High mass model} with $M_1=170 \rmModot$ and $M_2=80 \rmModot$ (\citealt{KashiSoker2010}, where the model was referred to as the `MTz model'; \citealt{KashiSoker2015, Kashi2017, Kashi2019}).
\end{enumerate}
We shall mark runs of the conventional mass model with `C', and for the high mass model with `M'.
Before the GE the mass of the primary was larger by $\approx 20$--$40  \rmModot$, for both models, and the mass of the secondary may have been $\approx 4 \rmModot$ larger for the high-mass model.

Presently, the orbital period is $5.54 \yr$, implying the semi-major axis is $a=16.64 \au$ for the conventional mass model, and $a=19.73 \au$ for the high mass model.
As explained above, at the time of the GE, however, the pre-GE orbital period was only $\simeq 5.1 \yr$, considerable amount of mass was lost from the system, and mass was transferred between the stars. All these effects combined suggest that the semi-major axis varied during the GE (see figure 5 in \citealt{KashiSoker2010}).
The average values of the semi-major axis during the GE are $\left< a \right> \simeq 16 \au$ for the conventional mass model, and $\left< a \right>  \simeq 19.75 \au$ for the high mass model.
Though the secondary receives and ejects mass, we do not modify its position according to the deviation from the Keplerian orbit during the periastron passage. 

The ambient gas profile is
\begin{equation}
\begin{split}
\rho_{w}(r) &= \frac{\dot{M}_w}{4 \pi r^2 v} \\
&= 5.5 \times 10^{-9}
\left(\frac{\dot{M}_w}{1 \msyr}\right)
\left(\frac{v_w}{500 \kms}\right)^{-1}\\
&\times
\left(\frac{r}{60 \rmRodot}\right)^{-2} \g \cm^{-3}, 
\end{split}
\label{eq:rhow1}
\end{equation}
where $v_w$ is the primary wind velocity and $\dot{M}_w$ is the mass loss rate in the primary wind.

We assume, as in earlier papers \citep{Soker2001,KashiSoker2010,KashiSoker2015}, that the secondary accretes gas ejected from the primary during the eruption, and that the gas forms an accretion disk (probably a thick disk, as the accretion time is short compared to the cooling time and viscous time).
Furthermore, we assume that the accretion disk launches two opposite jets. Our simulations follow the interaction of these jets with the ambient gas.

The jets are not expected to form at the moment accretion begins, but rather after some time when enough gas with sufficient angular momentum has been accumulated to form an accretion disk, or a thick accretion belt \citep{KashiSoker2009b}.
As simulations \citep{Kashi2017,Kashi2019} show that accretion predates the time of periastron passage, we take the duration of jet launching to start 2 days before periastron passage, for a duration of 20 days.

Each of the two jets is launched from the position of the secondary at a semi-opening angle of $\theta_{\rm jet}$.
The symmetry axis of the jet cones is identical to the axis connecting the poles of the secondary (the $z$ axis), namely perpendicular to the orbital plane (the $x$--$y$ plane).
The mass loss rate of the jets is $\dot{M}_{\rm jet} = 0.12$--$1.2 \msyr$, and they are ejected at velocity equal to the wind velocity of the secondary, $v_{\rm jet} = 3\,000 \kms$.
For these values of $\dot{M}_{\rm jet}$, the total mass ejected by jet in one periastron passage is $M_{\rm jet} \simeq 6.5 \times 10^{-3}$--$0.065 \rmModot$.
This is a small fraction of the mass available for the secondary to dispose of every periastron passage.

To solve the hydrodynamic equations we use the \textsc{flash} version of the split piecewise parabolic method (PPM) solver \citep{ColellaWoodward1984}.
The code includes radiative cooling based on \cite{SutherlandDopita1993}.
We use up to 8 levels of refinement with adaptive mesh refinement (AMR).
The grid is refined according to the distribution of density and temperature, using the standard \textsc{paramesh} refinement condition that is based on L\"ohner  error estimator (formulae are available in the \textsc{flash} manual\footnote{\url{http://flash.uchicago.edu/site/flashcode/user_support}}, chapter 8.6.3).
The refine and de-refine cutoff parameter values are 0.8 and 0.2, respectively, and the filtering parameter is $\varepsilon=0.2$, so practically the high resolution maintains wherever it was previously required to be so.
At this resolution the smallest cell size is $1.465 \times10^{12}\cm$, so not only the fine details of the outflow are revealed, but even the secondary star itself is resolved.
The high resolution allows to follow in detail the gas as in the launch region of the jet and allows following the interaction of the jet with the ambient gas with the ability to resolve fine details.
The importance of using high resolution in wind simulations was emphasized in \cite{Kashi2017}, where it was demonstrated that low resolution may fail to resolve details in the structure of the gas, such as clumps and filaments, and prevents instabilities to be triggered.

Table \ref{table:parameters} presents a list of the simulations we ran and the parameters we used.

\section{RESULTS}
\label{sec:results}

We start with run C1, which simulates the conventional mass model, and has narrow jets with a semi-opening angle $\theta_{\rm jet} = 15^\circ$ and mass loss rate $\dot{M}_{\rm jet} = 0.6 \msyr$ (combined and divided equally between the two opposite jets). The density profile of the ambient gas as in equation \ref{eq:rhow1}.

\begin{figure*}
\centering
\includegraphics[width=0.75\textwidth]{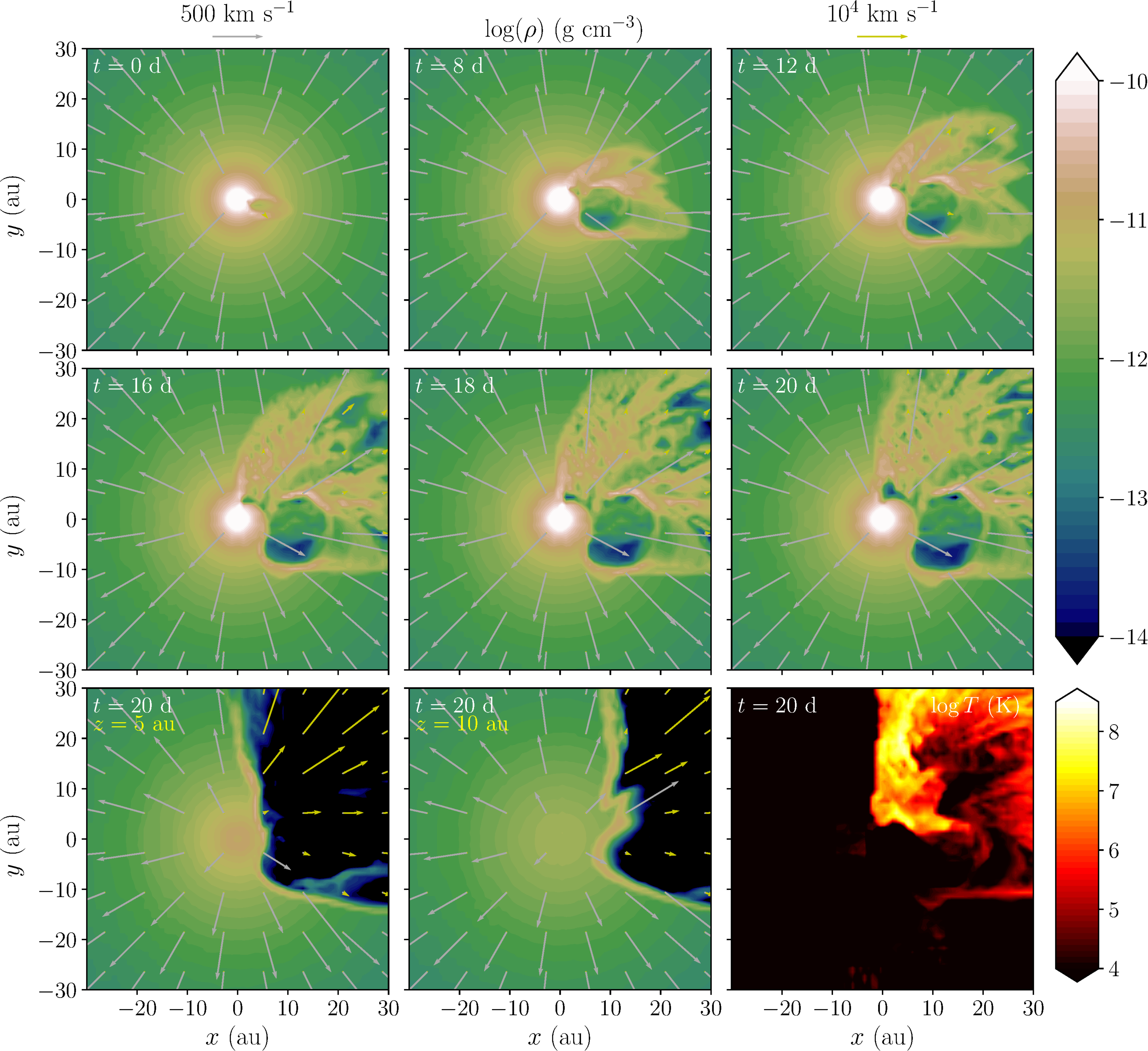}  
\caption{
Density maps showing the density slices for run C1 (conventional mass model).
The upper and middle row shows the density in the orbital plane ($z=0$), at different times relative to periastron passage ($t=0$).
The secondary is moving counterclockwise with time, being to the right of the primary at ($t=0$).
The colorbar on the upper right indicates $\log \rho$.
The left and middle panels on the bottom row show higher latitude slices ($z=5\au$ and $z=10\au$, respectively) for the same time as in the right panel on the middle row.
The right panel on the bottom row shows a temperature map for the same slice as in the right panel on the middle row, with the colorbar on the bottom right indicating $\log T$.
Vectors indicate 2d velocity on the sliced plane.
Note the two scales for velocity vectors (indicated by arrows on the top of the figure), created to more easily distinguish between the fast and slow flow.
}
\label{fig:dens_C1}
\end{figure*}

A density map showing a slice on the orbital plane of run C1 is presented in Figure \ref{fig:dens_C1}. The last panel in this figure shows a temperature map. 
We can see that the lobes have low density and very hot gas.
In the simulation the jets are ejected perpendicular to the orbital plane, namely in the $\pm z$ axis. However, interaction with the ambient outflowing gas (primary wind) accelerates the latter to high velocities and diverts the gas to lower latitudes.

\begin{figure*}
\centering
\includegraphics[trim= 4.1cm 0.0cm 1.1cm 1.5cm, clip=true,width=0.99\textwidth]{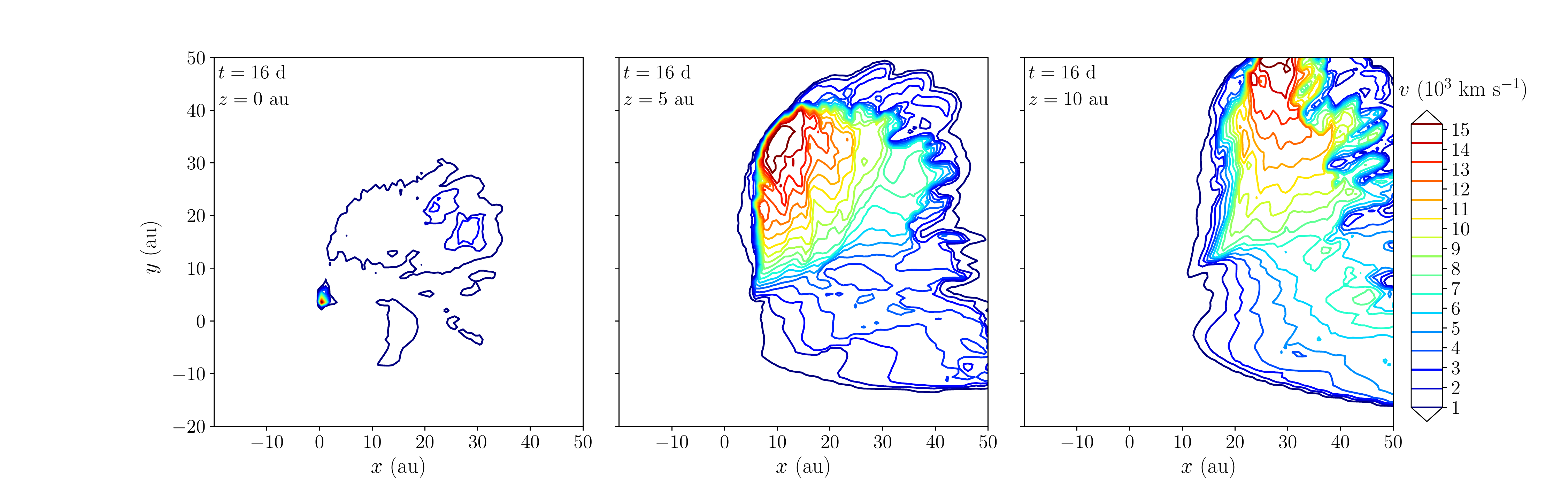}  
\caption{
Velocity contour maps showing at slices for run C1, 16 days after periastron passage.
The left, middle and right panels show latitude slices for $z=0\au$, $z=5\au$ and $z=10\au$, respectively).
Note that the figure is not centered on the primary (which is at $(x=0,y=0)$ on the left panel), but rather shows the extensive region of the accelerated gas.
Even faster moving gas than the maximum value seen in this figure exists at this run, but not at those specific slices.
}
\label{fig:vel_C1}
\end{figure*}

A very fast outflow is developed as a result of the interaction between the jets and the ambient gas of the GE.
The flow is at first towards the point where the secondary is at periastron (orientation angle $\omega=0$, measured from the line connection the primary and the point of periastron passage), and later towards larger values of $\omega$.
A better view of the velocities is shown in Figure \ref{fig:vel_C1}, with contours emphasizing the regions where the velocities are high.
The time presented in Figure \ref{fig:vel_C1} is the time when $M(>10^4 \kms)$ is maximal.
%
%

\begin{figure*}
\centering
\includegraphics[trim= 0.0cm 0.0cm 0.0cm 0.0cm, clip=true,width=0.8\textwidth]{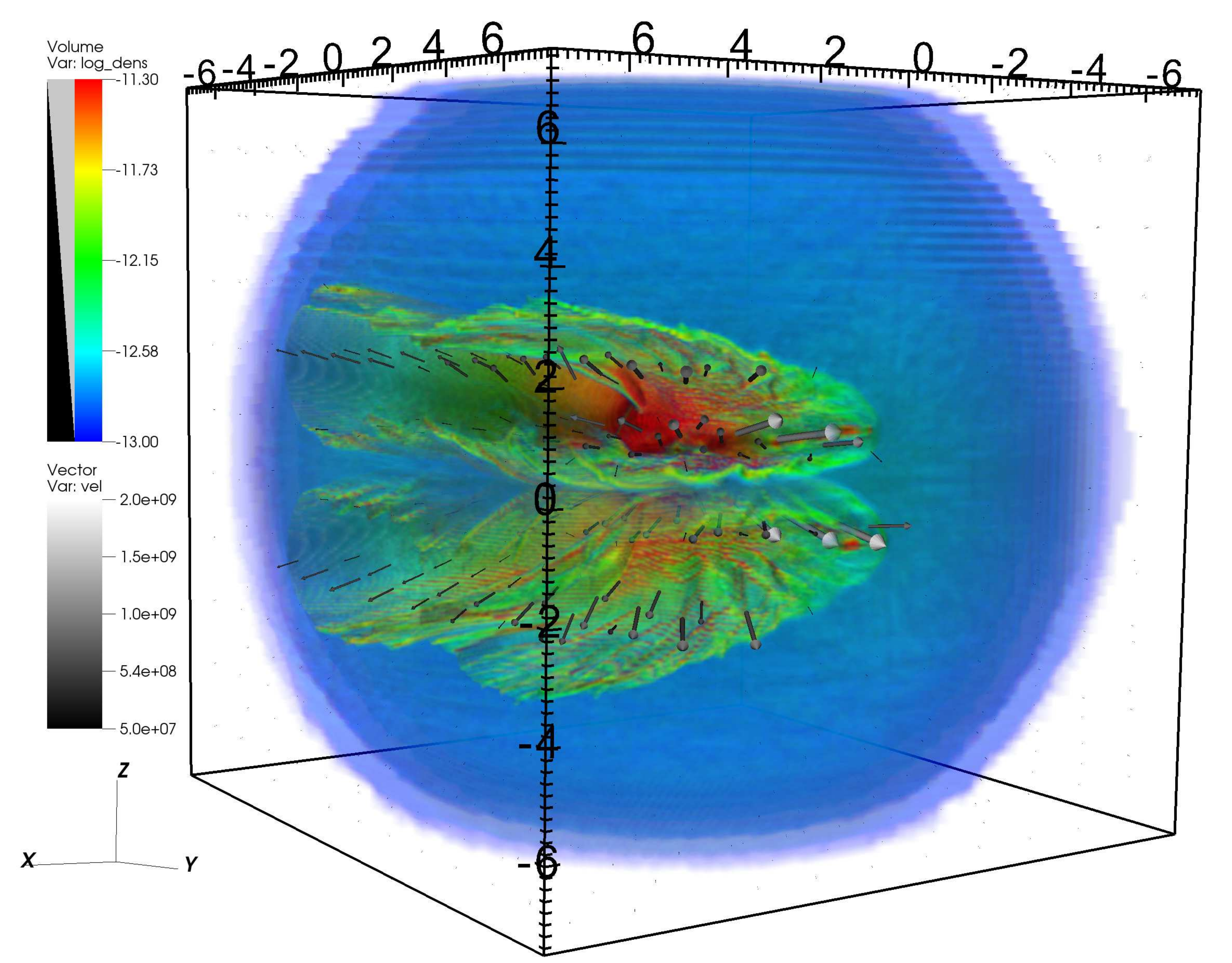}  
\caption{
A three-dimensional view of the high velocity mass in run C1 (conventional mass model), taken 12 days after periastron passage. Axes units are $10^{14} \cm$ (each side of the box is $100 \au$.) The jets are launched at $3\,000 \kms$ in the $\pm z$ directions, perpendicular to the orbital plane.
The blue sphere marks where the primary wind reaches $10^{-13} \,\rm{g}\,\rm{cm^{-3}}$.
The arrow size is linearly scaled between $500 \kms$ and $20\,000 \kms$.
The flow is deflected away from polar directions, creating a two-lobed low density cavity in the wind ejected from the primary. This is the place where we find the high velocity gas with $v \ge 10\,000 \kms$.
The density is largest and the velocities are highest in the direction closer to the line connecting the primary and the secondary.
}
\label{fig:3d}
\end{figure*}

In Figure \ref{fig:3d} we present a 3D view of the mass that expands at high velocities.
We can see that the flow creates a lower density and high velocity cavity in the primary wind. The velocity direction points approximately from the primary to the secondary, and the small deviation from this direction is a result of the secondary orbital velocity.

\begin{figure*}
\centering
\includegraphics[trim= 1.0cm 0.15cm 2.05cm 1.55cm, clip=true,width=0.7\textwidth]{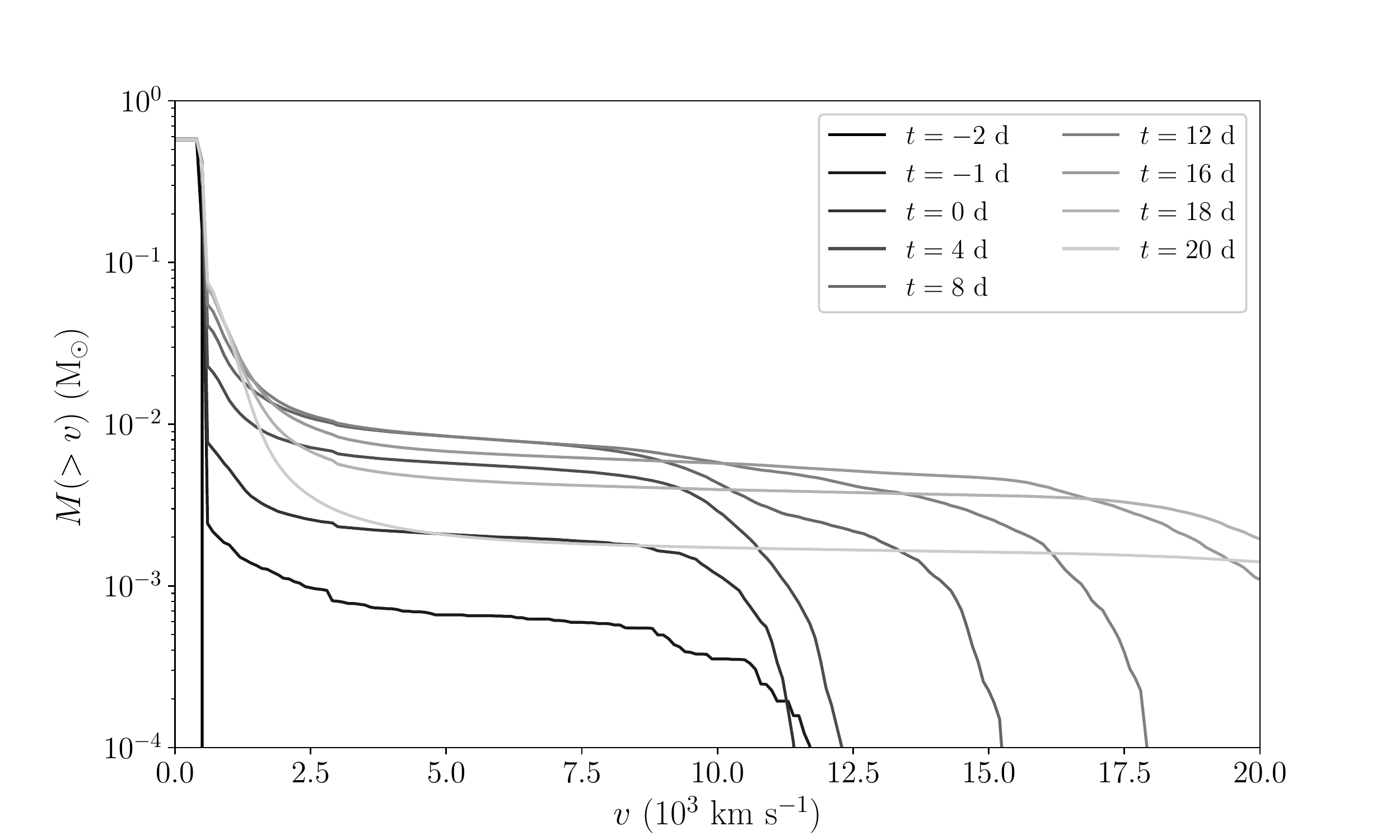}
\caption{
The gas mass at velocity above $v$, for run C1.
Grey scale lines from dark grey to bright grey correspond to the velocity profile at different time with respect to periastron passage, as indicated in the legend. The flow produces velocities larger than $10^4 \kms$. As the secondary moves across periastron passage more material reaches high velocities.
As time progresses more material is being accelerated into high velocities.
The jets are only injected for 20 days, close to periastron passage (from day $-2$ till day $+18$, relative to periastron passage). After that, the amount of fast gas moving at high velocities decreases.
}
\label{fig:mass_vel_C1}
\end{figure*}

\begin{figure*}
\centering
\includegraphics[trim= 1.0cm 0.15cm 2.05cm 1.55cm, clip=true,width=0.7\textwidth]{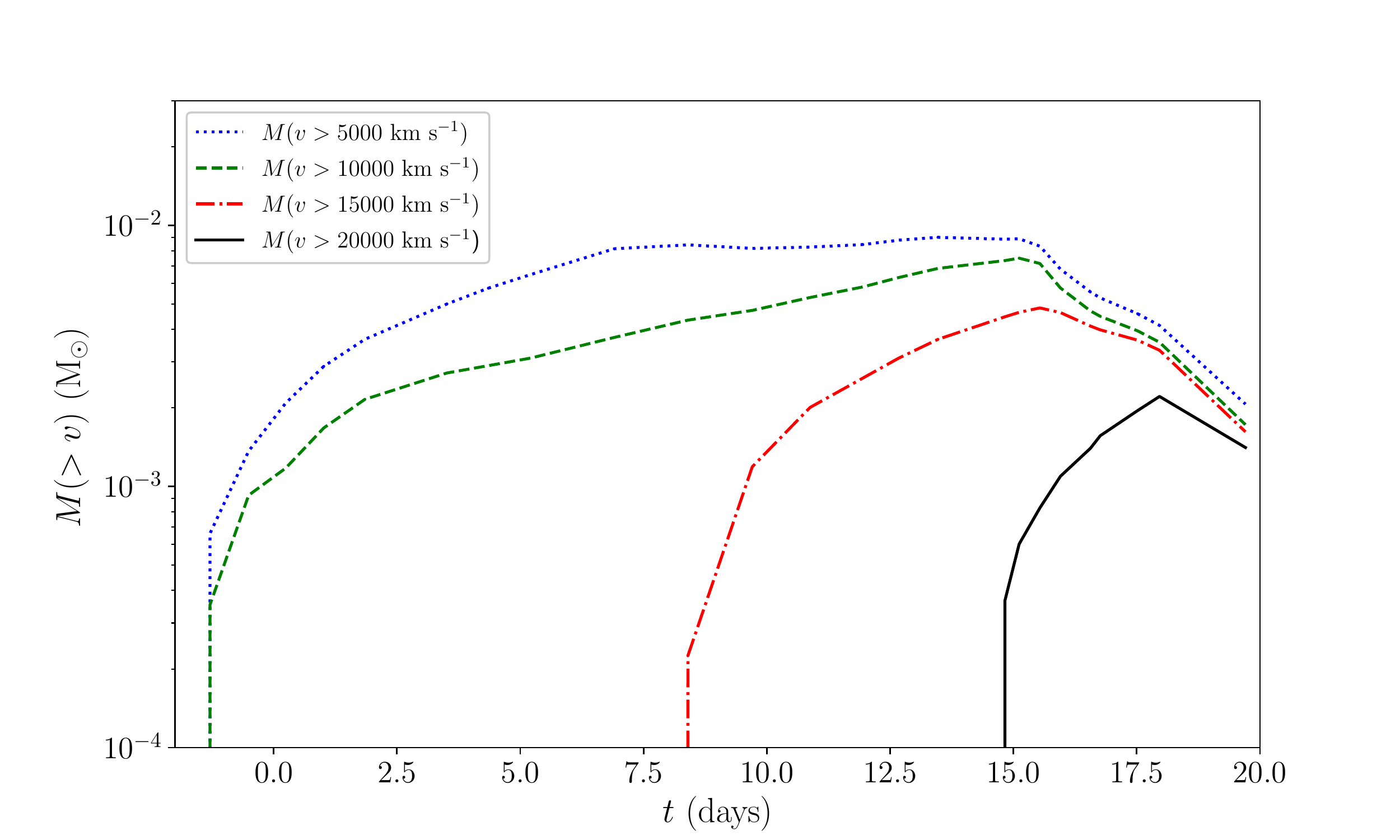}
\caption{
The gas mass at velocity above $v$, for run C1, as a function of time.
Four velocity bins are taken, as indicated in the legend.
We can see that the fast velocity mass reaches a maximum, at $t\simeq 16 \days$. 
}
\label{fig:mass_vel_t_C1}
\end{figure*}

We post-processed the simulation to measure $M(>v)$, the mass of the gas traveling at velocity above $v$.
Figure \ref{fig:mass_vel_C1} shows that a fraction of the mass indeed travels at velocity larger than $10\,000 \kms$, at different times during the simulation.
The values of $M(>10^4 \kms)$ listed in Table \ref{table:parameters} are for time they are at maximum.
The velocity distribution is time dependent.
Figure \ref{fig:mass_vel_t_C1} presents $M(>v)$ as a function of time. From these two figures we can follow the gas-jet interaction. More and more mass reaches higher and higher velocities, until a maximum is reached. The higher the velocity, the later the time it reaches maximum. As an indicator, $M(v>10^3 \kms)$ reaches a maximum at $t\simeq 16 \days$.
After $t = 18 \days$ there is a decline in all the velocity bins, as the launching of the jets has stopped.
At later times the fast velocity gas continues to collide with the ambient primary wind, and transfers the kinetic energy to the slower gas.
%
%

The orbital period at the time of the GE was not constant during the GE. As shown in \cite{KashiSoker2010},
the period increased from $\simeq 5.1 \yr$  to $\simeq 5.5 \yr$ during the $\simeq 20 \yr$ duration of the GE, as a result of the combined effect of mass ejection from the binary system and mass transfer from the primary to the secondary.
The GE lasted about four revolutions, namely four periastron passages.
During this entire time wind was ejected from the primary.
The accretion in present day \etc occurs near periastron passage, due to the very high eccentricity of the system.
We assume that most of the accretion during the GE also occurred near periastron passages, though it surely extended for longer periods at weaker rates, maybe even throughout the entire orbit.

In our simulations we only follow a fraction of the orbital trajectory, close to periastron passage.
It would have been ideal to simulate the duration of the entire GE, but unfortunately this is presently not possible with the available computational resources at our disposal.
Nevertheless, much can be learned on the GE from the results of one periastron passage, in light of the accretion model.

Therefore, the correction we will use for the total mass of the fast-flowing material will be obtained by multiplying the results by 4, the number of periastron passages.
Taking the results of run C1, where we obtained that the mass reaching high velocity during one periastron passage is $M_{h,1} \approx M(>10^4 \kms) \simeq 0.0058 \rmModot$, we can estimate the mass ejected for the entire GE is $M_h \approx 0.023 \rmModot$.
Namely, we find that the value of $M_h$ based on our simulation is the same order of magnitude as the value we earlier derived from theoretical considerations based on the accretion model for the GE and the observations of \cite{Smithetal2018a}.

\begin{figure*}
\centering
\includegraphics[trim= 2.0cm -0.3cm 1.2cm 1.3cm, clip=true,width=0.99\textwidth]{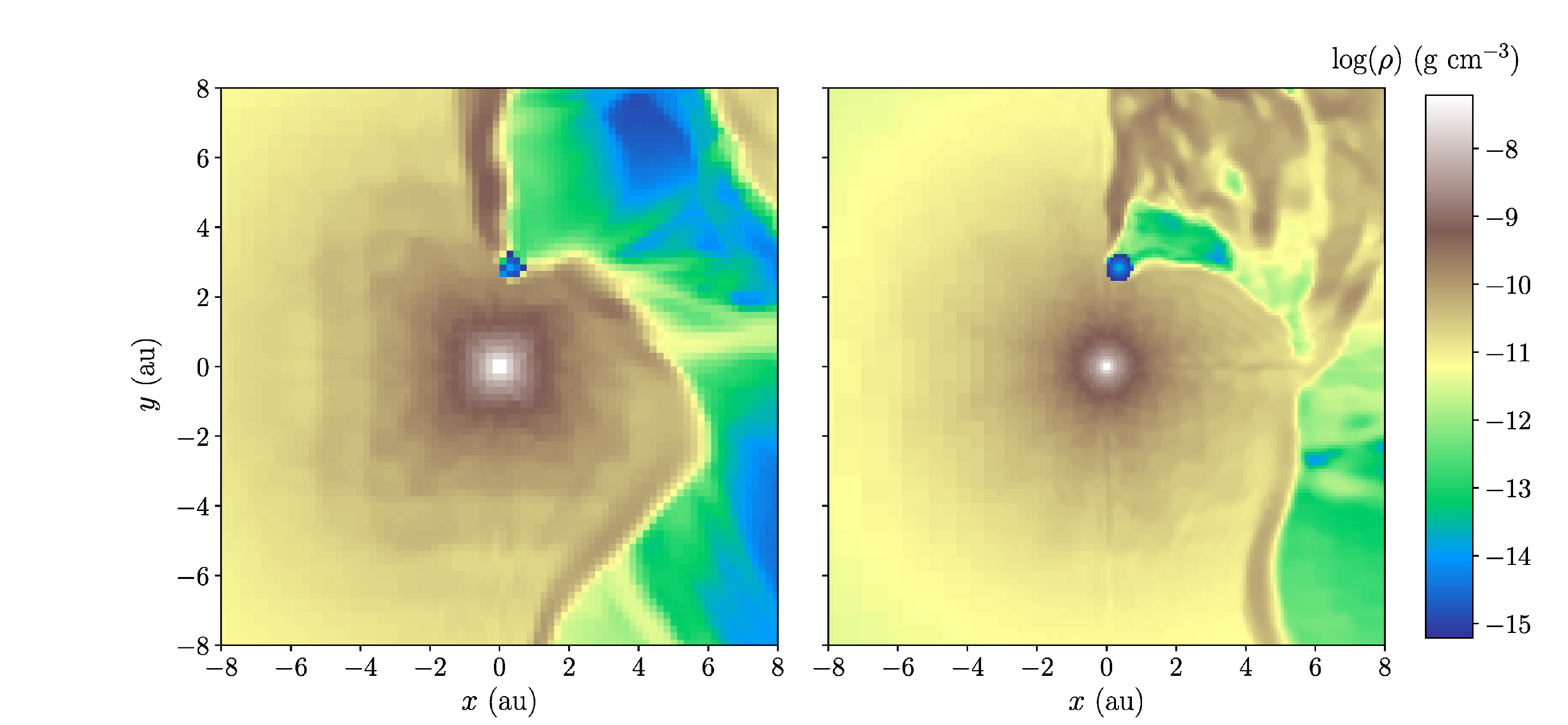}
\caption{
\textit{Left}: A density map of a run similar to C1, but with 7 levels of refinement instead of 8.
\textit{Right}: A density map of run C1.
Both panels zoom on the central part of the grid, on the plane $z\simeq0$, and at $t= 16 \days$ after periastron passage.
The primary is at the center and the secondary is above.
The differences between the two panels are apparent, especially at the ability to resolve the jet flow close to the secondary and the ability to obtain spherically symmetric primary wind.
}
\label{fig:res_zoom}
\end{figure*}

Figure \ref{fig:res_zoom} demonstrates the importance of high resolution.
It focuses on the central part of the grid, close to the binary, presenting run C1 on the right hand side, and the same run with lower spatial resolution (7 levels of refinement, namely resolves spatial scale twice the size of C1) on the left hand side.
The differences are considerable, and for the higher resolution simulation we can see that (1) the shape of the wind is much closer to being spherically symmetric; (2) the jet outflow is not degraded and is closer to being ejected as a cone (looks very close to circle in the slice that is taken just one cell above $z=0$); (3) the interaction between the jet and the primary wind and consequently is better resolved and produces more accurate hydrodynamic properties.
%

\begin{figure*}
\centering
\includegraphics[trim= 1.0cm 0.15cm 2.05cm 1.55cm, clip=true,width=0.7\textwidth]{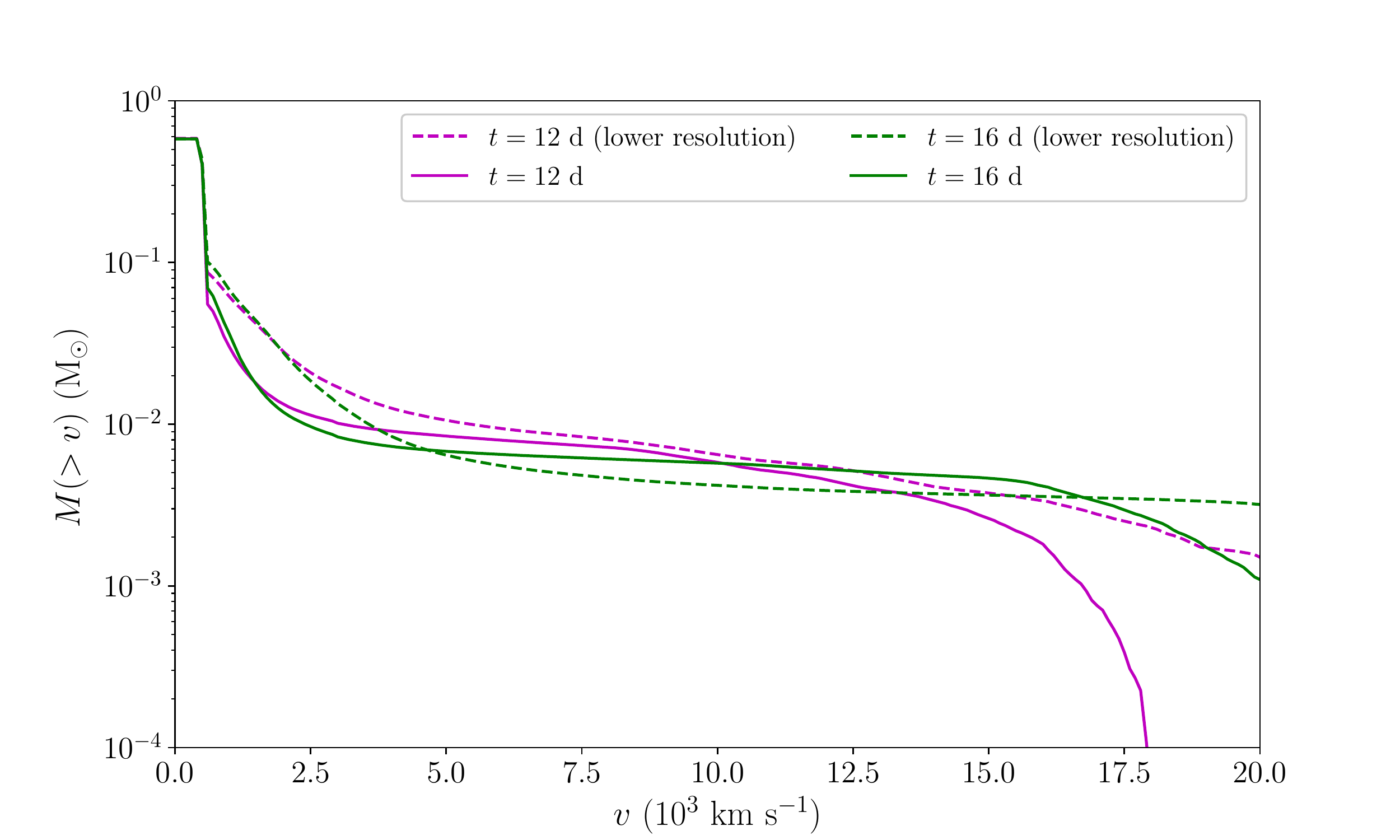}
\caption{
The gas mass at velocity above $v$, for run C1 and a similar run with lower resolution.
Despite the difference in the flow, which is more considerable close to the secondary, the curves have similar shapes and the mass is of the same order of magnitude up to about $ 17\,000 \kms$. Namely, the effect of of fast jets accelerating ambient wind can be seen also in the lower resolution, though it produces less accurate results.
}
\label{fig:res_compare_v}
\end{figure*}

We post-processed the lower resolution simulation, which has similar parameters as run C1 but 7 levels of refinement instead of 8.
Figure \ref{fig:res_compare_v} shows a comparison of the results of the gas mass at velocity above $v$.
While there are differences between the flows in the two inspected resolutions, the overall effect in accelerating the primary wind to very high velocities is seen in both.
For the lower resolution simulation, the value of $M(v > 10^4 \kms)$ is $\simeq 10$ per cent larger at $t= 12 \days$, but about $\simeq 2/3$ at $t= 16 \days$.
The amount of mass at velocities $\gtrsim 17\,000 \kms$ is consistently larger for the lower resolution simulation, and therefore the lobes expand faster. The differences originate mainly close to the secondary where the jets are ejected and the acceleration of the ambient gas begins.
Overall, the effect of jets accelerating the primary wind to high velocities appears even in the low resolution, and 
the amount of mass accelerating to high velocities is quite close.
%

After describing the fiducial simulation C1, we now turn to the other runs we examined. Of course, the parameter set is large and we only focused on a small sub-set of parameters.
The runs we selected (C2--C4) are extreme cases. They were selected so that if they all show that mass is accelerated to high velocities, we will safely be able to conclude that this is the case for the wide range of those parameters.

\begin{figure*}
\centering
\includegraphics[width=0.75\textwidth]{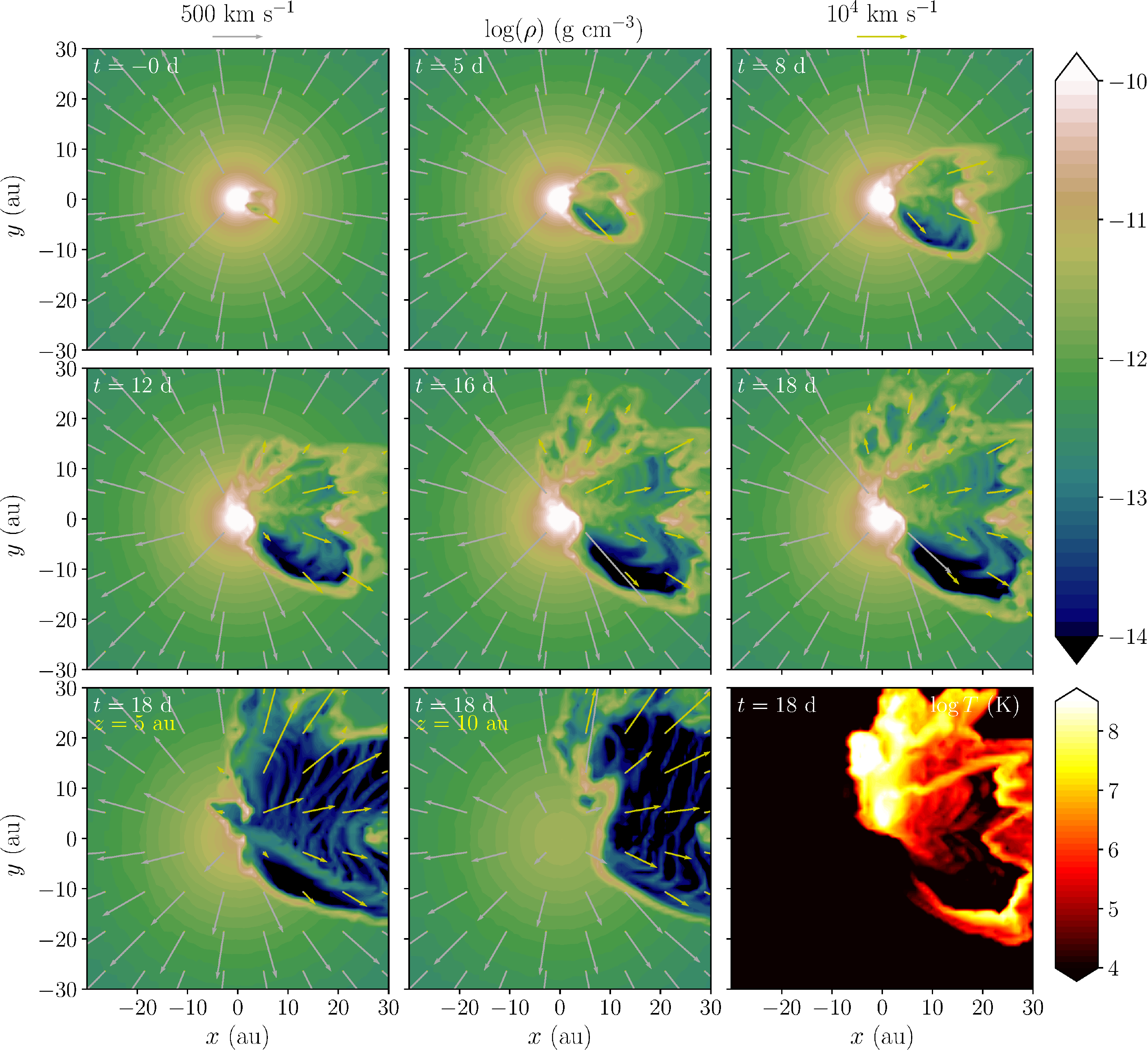}  
\caption{
Same as Figure \ref{fig:dens_C1}, but for run C2.
}
\label{fig:dens_C2}
\end{figure*}

\begin{figure*}
\centering
\includegraphics[trim= 1.0cm 0.15cm 2.05cm 1.55cm, clip=true,width=0.75\textwidth]
{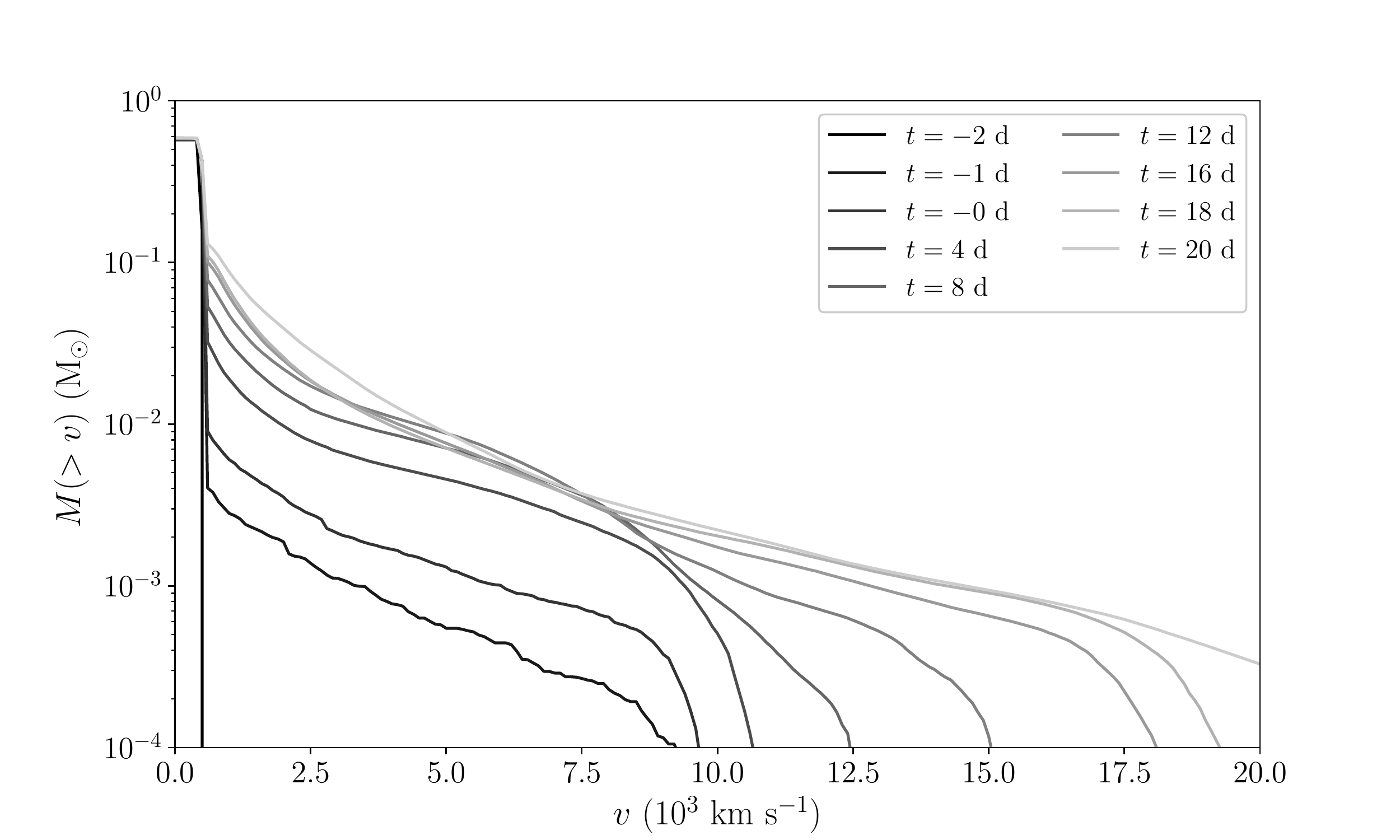}
\caption{
The gas mass at velocity above $v$, for run C2.
}
\label{fig:mass_vel_C2}
\end{figure*}

The first parameter we checked is the width of the jets, as this parameter was not motivated from any physical consideration.
The value we used in run C1 is quite a narrow jet.
But jets are not necessarily narrow, and it is possible to have very wide jets, e.g., as observed by \cite{Bollenetal2019} in a post asymptotic giant branch binary system.
Run C2 is the same as C1 except that the jet (or collimated outflow) is much wider, with semi-opening angle of $70^\circ$.
Figures \ref{fig:dens_C2} and \ref{fig:mass_vel_C2} show the results for this run.
A wider jet, even though having the same energy as a narrow jet, has lower density. It is therefore expected that it will
have a weaker interaction with the ambient primary wind than the narrow dens jet. 
As expected, the amount of high velocity gas that is obtained, is $\simeq 2.1 \times 10^{-3} \rmModot$, about a third of that of run C1 (which has a narrower jet).
Even though the jets in run C2 are wider, the opening in the primary wind is still in the equatorial periastron direction.
The conclusion is that the jet is able to deposit its kinetic energy in the primary wind almost independent on its opening angle.
%
%

\begin{figure*}
\centering
\includegraphics[width=0.75\textwidth]{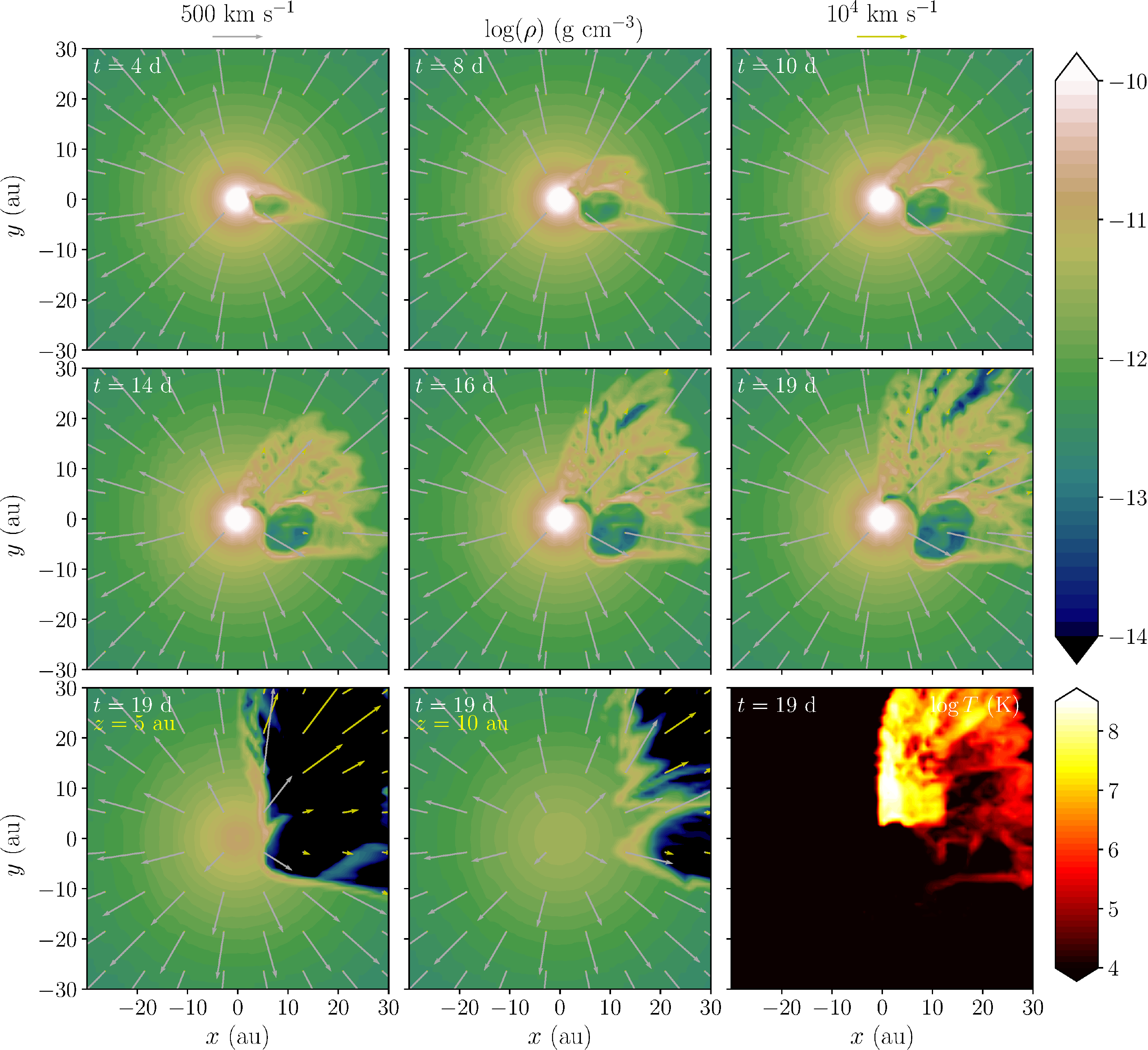}  
\caption{
Same as Figure \ref{fig:dens_C1}, but for run C3.
}
\label{fig:dens_C3}
\end{figure*}

\begin{figure*}
\centering
\includegraphics[trim= 1.0cm 0.15cm 2.05cm 1.55cm, clip=true,width=0.7\textwidth]{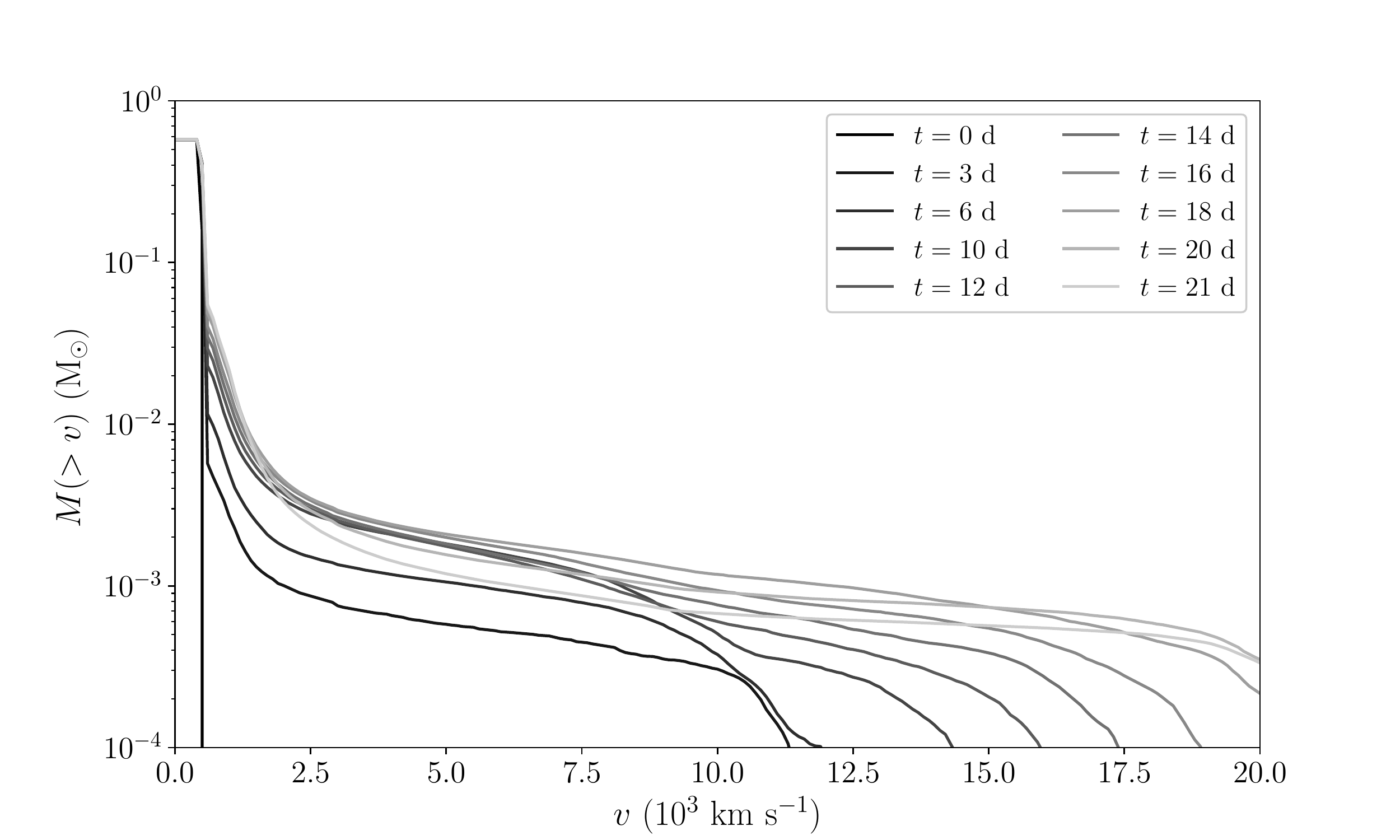}
\caption{
The gas mass at velocity above $v$, for run C3.
}
\label{fig:mass_vel_C3}
\end{figure*}

\begin{figure*}
\centering
\includegraphics[width=0.75\textwidth]{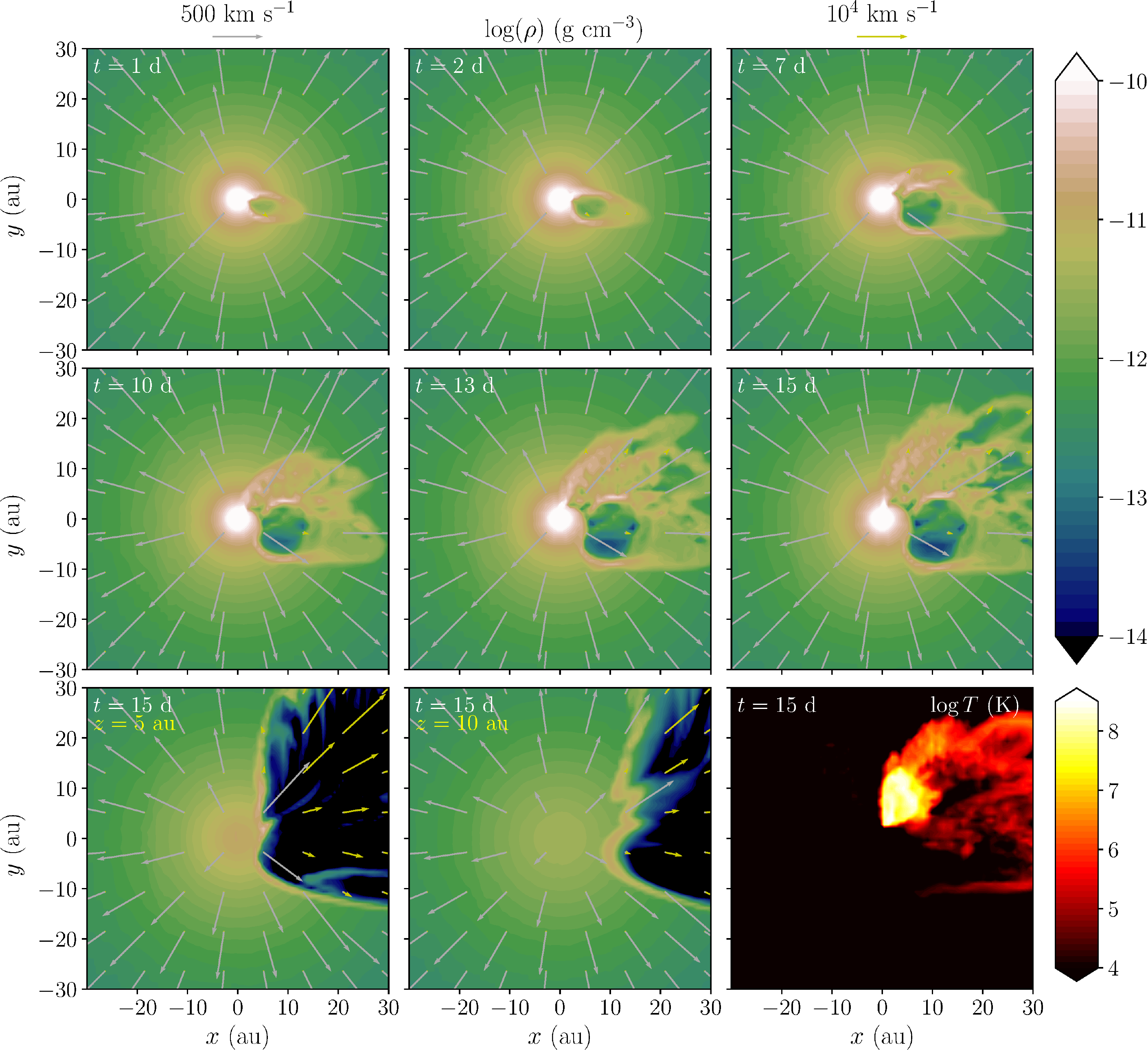}  
\caption{
Same as Figure \ref{fig:dens_C1}, but for run C4.
}
\label{fig:dens_C4}
\end{figure*}

\begin{figure*}
\centering
\includegraphics[trim= 1.0cm 0.15cm 2.05cm 1.55cm, clip=true,width=0.7\textwidth]{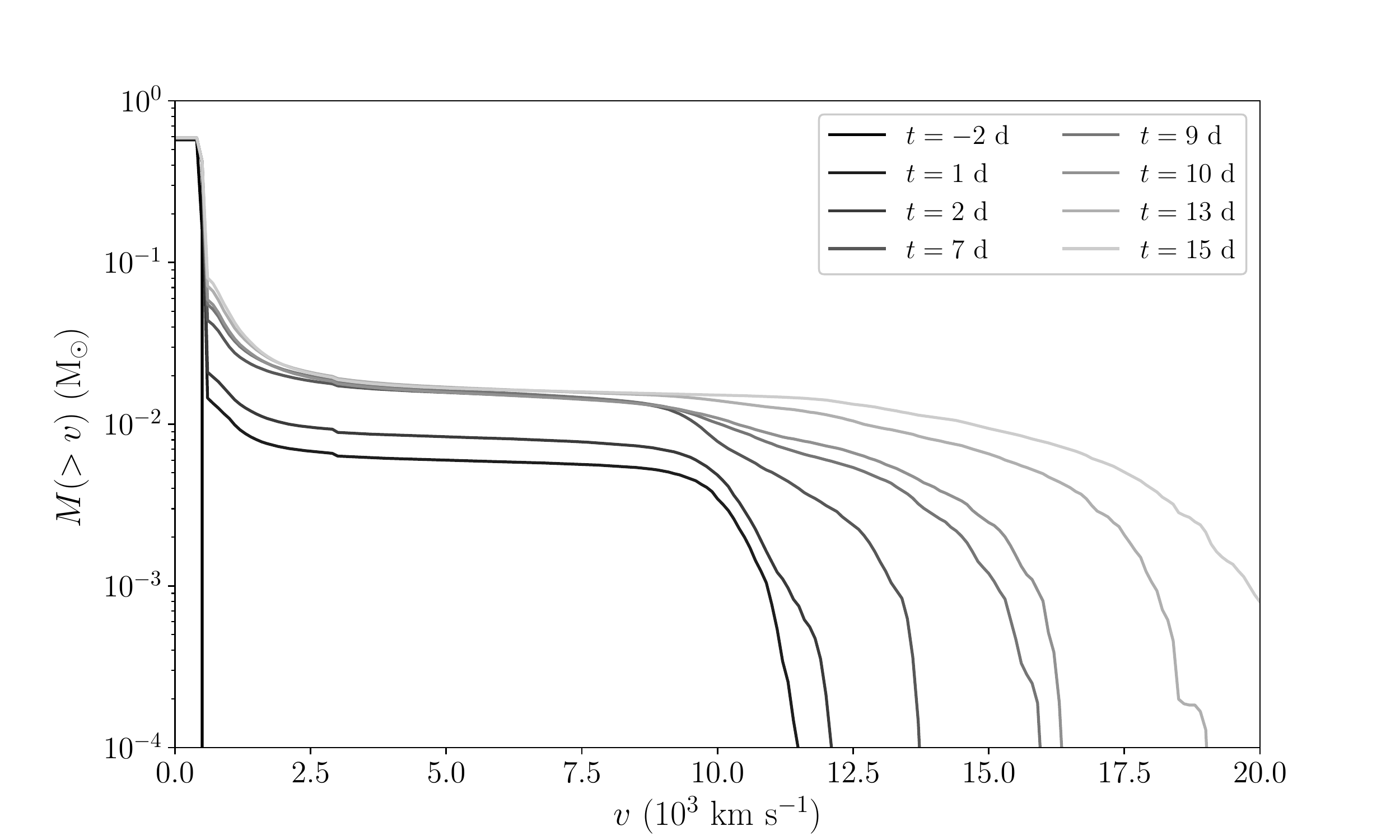}
\caption{
The gas mass at velocity above $v$, for run C4.
}
\label{fig:mass_vel_C4}
\end{figure*}

Next, we examined run C3, where the mass loss rate of the jets is much smaller $\dot{M}_{\rm jet} = 0.12 \msyr$.
The results are shown in Figure \ref{fig:dens_C3} and Figure \ref{fig:mass_vel_C3}.
The time it takes the jets to penetrate the primary wind is longer.
We see that the amount of mass accelerated to high velocities is also much smaller, $\simeq 1.2 \times 10^{-3} \rmRodot$.
We find the relation to be linear.
Namely, jets ejecting fifth of the mass accelerate about fifth the amount of mass to high velocities.
On the other extreme, we examined run C4, where the mass loss rate of the jets is much larger $\dot{M}_{\rm jet} = 1.2 \msyr$.
The results are shown in Figure \ref{fig:dens_C4} and Figure \ref{fig:mass_vel_C4}.
There, we find the amount of mass accelerated to high velocities to be $1.6 \times 10^{-2}$.
This increase of the mass in the jets by a factor of five, resulted only an increase of the amount of high velocity gas by a factor of less than 3.
The reason is the fast and intense interaction of the jets and the primary wind, that resulted in some mass getting out of our grid, therefore unaccounted for.
The strong interaction also caused numerical complication that prevented the simulation from being completed to day 20 as the other simulations.
Under better simulation conditions we think the ratio would have stayed linear.
%
%
%
%
%

From runs C1--C4 we conclude that there is dependency on the parameters we varied, but it is not a dramatic one. Within the range of parameters we studied, the jets have proven to be able to accelerate substantial amount of gas to high velocities.
As mentioned, we are limited with the number of runs we can perform, but the conclusions from the ones we did is very encouraging.

\begin{figure*}
\centering
\includegraphics[width=0.75\textwidth]{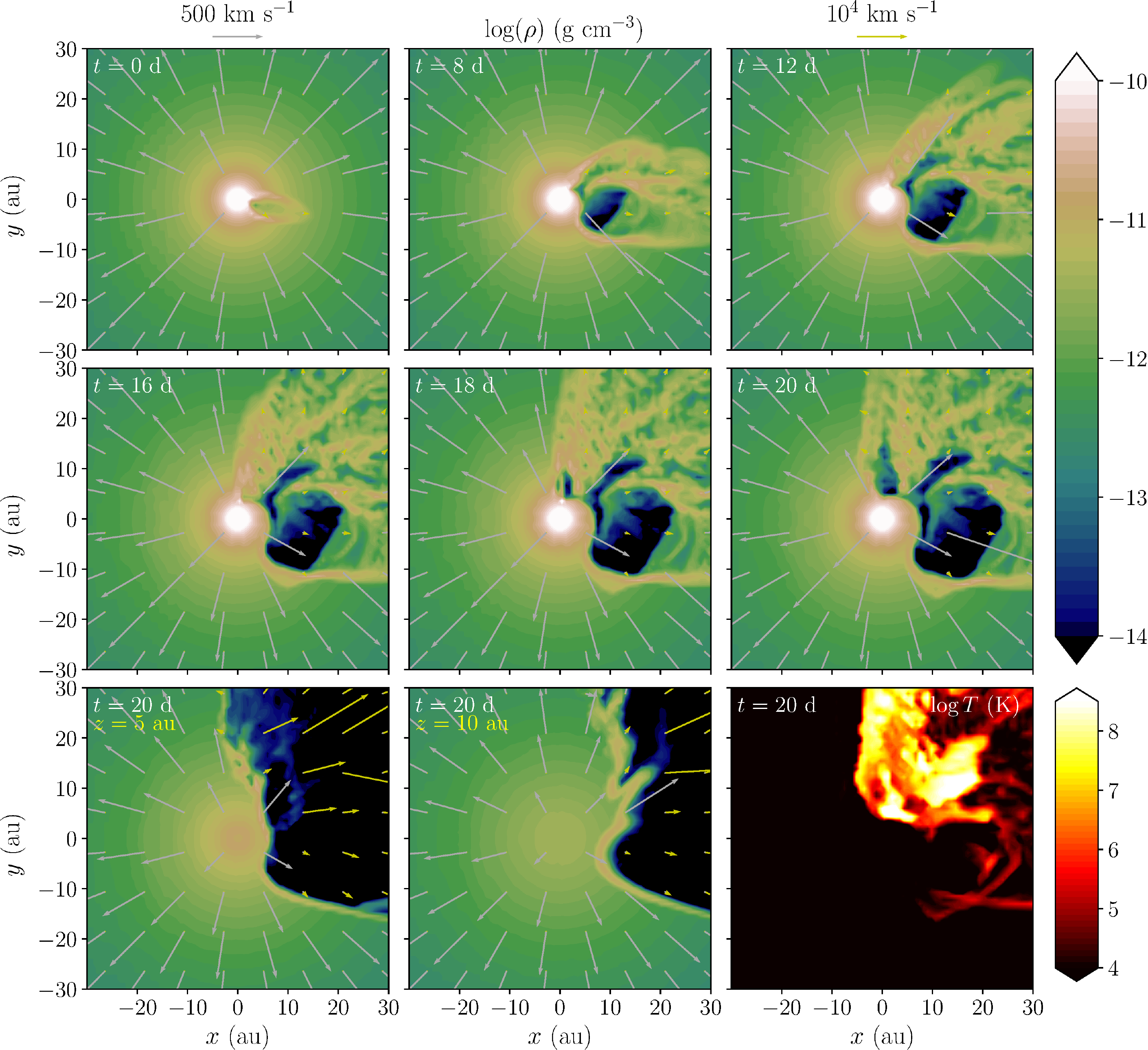}  
\caption{
Same as Figure \ref{fig:dens_C1}, but for run M1.
}
\label{fig:dens_M1}
\end{figure*}

\begin{figure*}
\centering
\includegraphics[trim= 1.0cm 0.15cm 2.05cm 1.55cm, clip=true,width=0.7\textwidth]{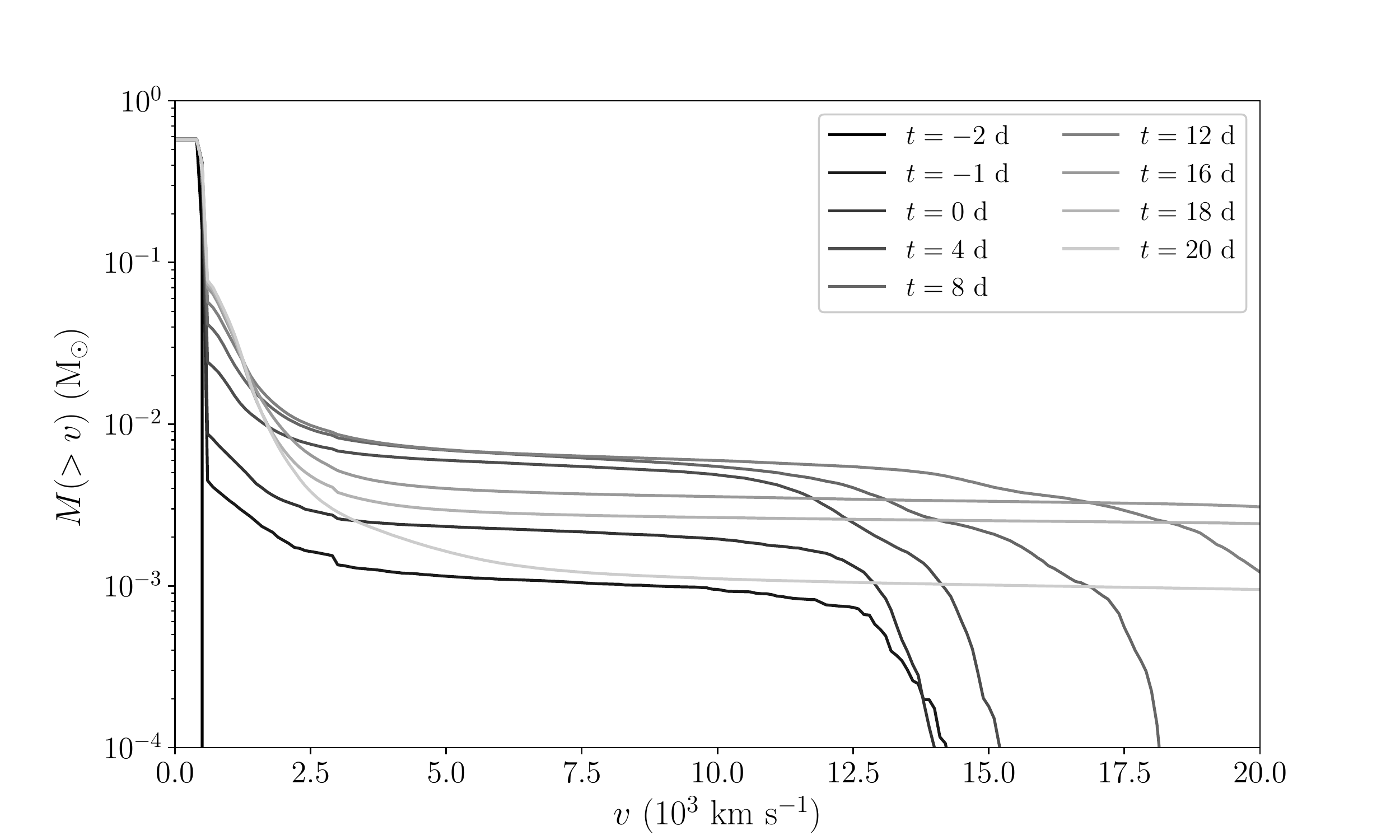}
\caption{
The gas mass at velocity above $v$, for run M1.
}
\label{fig:mass_vel_M1}
\end{figure*}

The most important run we now turn to examine, other than the fiducial run, is M1, which is similar to the fiducial rum C1, except that the orbit matches the one expected from the high-mass model of \etc.
The results are shown in Figure \ref{fig:dens_M1} and Figure \ref{fig:mass_vel_M1}.
The high velocity gas reaches a maximal value of $5.9 \times 10^{-3} \rmModot$. This mass is almost similar to the one obtained in run C1.
Figure \ref{fig:mass_vel_t_C1} shows that the maximum amount of mass traveling at high velocities, is obtained at earlier times compared to the conventional mass model, at day $\simeq + 14$ after pariastron passage.
This happens because the periastron distance is smaller for the high mass model, therefore the jets are launched in a denser region of the primary wind. The momentum exchange is therefore larger and the primary wind is accelerated.
%
%
%
\begin{figure*}
\centering
\includegraphics[trim= 1.0cm 0.15cm 2.05cm 1.55cm, clip=true,width=0.7\textwidth]{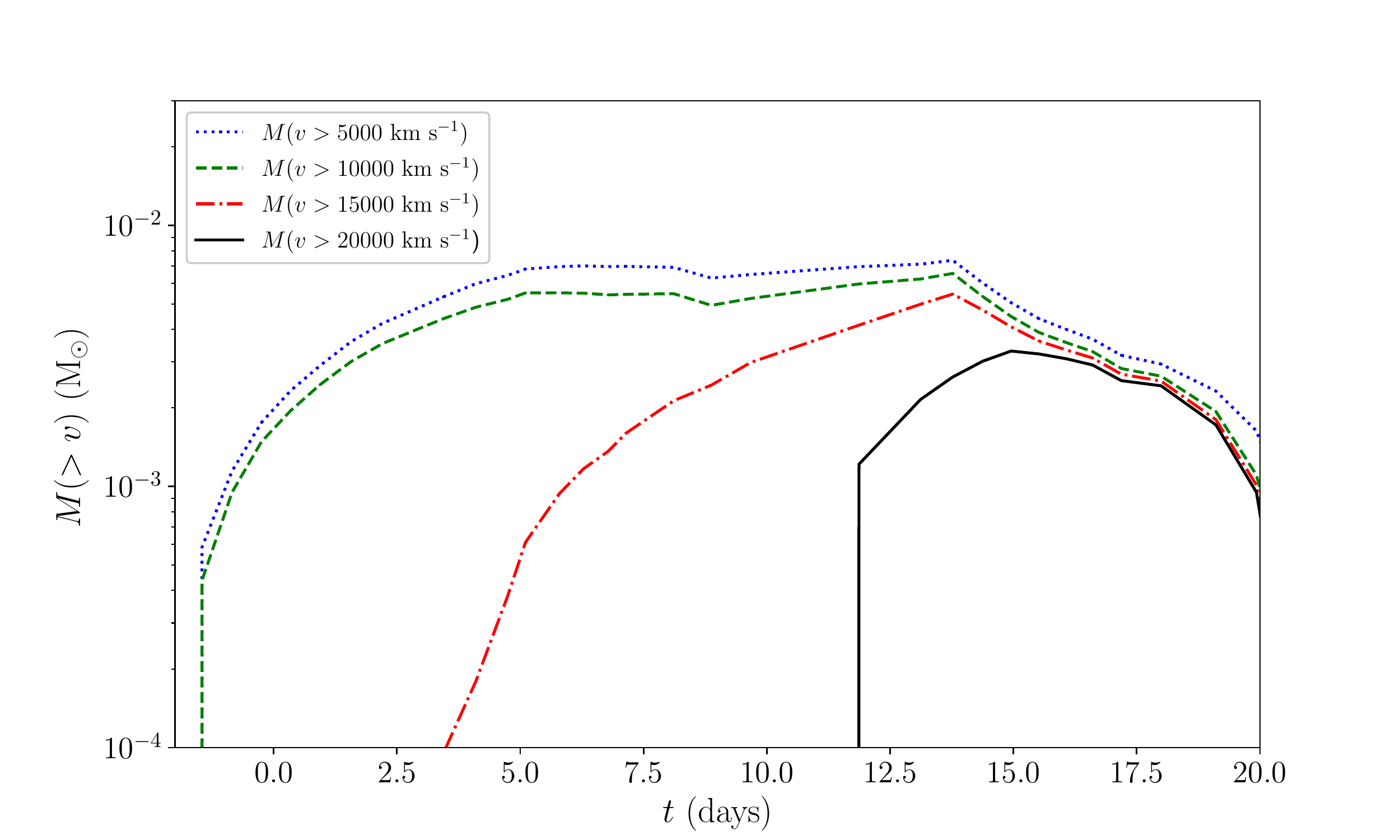}
\caption{
The gas mass at velocity above $v$, for run M1, as a function of time.
Four velocity bins are taken, as indicated in the legend.
We can see that the fast velocity mass reaches a maximum, at $t\simeq 13 \days$. 
}
\label{fig:mass_vel_t_M1}
\end{figure*}

\section{SUMMARY AND DISCUSSION}
\label{sec:summary}

By observing light echos, \cite{Smithetal2018a,Smithetal2018b} found that gas from the GE of \etc reached velocities of more than $10\,000 \kms$.
We have not studied before the possibility that material could reach these high velocities. Our simulations here are motivated by the new observations of \cite{Smithetal2018a,Smithetal2018b}.

Based on observations of giant eruptions and other transients that showed outflows at different velocities, \cite{Humphreysetal2016} suggested that
a single event such as a giant eruption can create multiple outflows as a result of material that flows in different velocities.
Moreover, they suggested that this may be a common phenomena.
Indeed, we see from the observations of \cite{Smithetal2018a} that the variance in velocities can be large.

We performed detailed 3D numerical hydrodynamic simulations of the accretion model for the GE of \etc, focusing on the jets ejected from the accretion disk and their interaction with the ambient gas.
We found that a fraction of the mass reaches high velocities of $> 10\,000 \kms$.
Our results, based on one periastron passage, corrected for the duration of the GE, imply that the amount of mass that reaches the high velocities is $M_h \approx 0.005$--$0.05 \rmModot$, depending on the selection of free parameters, with an intermediate value, $M_h \approx 0.02 \,\rm{M_{\odot}}$, more likely.
This value is time dependent and decreases when the secondary moves away from periastron and has less dense ambient gas in its vicinity with which the jets can interact.
Later, $M_h$ continues to decrease when the high velocity gas continues its interaction with the slow outflow of the primary and transfers its kinetic energy.

The high velocity gas may have been obscured at early times, and revealed only later times when it penetrated the much denser primary wind. By then, its amount may have been reduced as it interacted with the surrounding slower primary wind.

The results are in agreement with the assessment we made based on the observations of \cite{Smithetal2018a}, and the energy budget estimate of the accretion model (section \ref{sec:model}).
We also showed that the fast outflow can be obtained (1) for wide and narrow opening angle of the jets, (2) for high and low mass loss rate of the jets,
and (3) for both the high-mass and conventional-mass models of \etc.

The high mass model yields about the same amount of high velocity gas, but it is reaching its maximum a few days earlier than for the conventional mass-model.
Given the wide range of unknowns, it is at this point not possible to use light echo observations to make a distinction between the results of the two models, and favor one of them.
Nevertheless, the simulation results

Even though the flow looks somewhat different for the runs we studied, having very high velocity gas for all studies cases suggests that the accretion model for the GE, and the jets ejection it involves can account for the echo-observed high velocity gas of the GE. 

\cite{Smithetal2018a} commented that as observations were taken from an echo that views \etc
from the equator, these high speeds are probably not indicative of a polar jet, not even one from a compact object.
They further commented that the high velocities in echoes seen from the equator combined with fast polar velocities in the outer ejecta seen today \citep{Smith2008} suggest a wide-angle explosion rather than a highly collimated jet.

The jets we simulated here were launched in polar directions at velocity of $3\,000 \kms$. As a result of interaction with the ambient gas, the latter was accelerated to high velocities, so that the fast flowing material that reached velocities $> 10\,000 \kms$ came from low latitude directions,
Obtaining this gas at high velocities matches the direction of the echos from where \cite{Smithetal2018a} observed their spectra.

Jets are though to have a major impact in many processes where binaries interact. They have been suggested to explain intermediate-luminosity optical transients (ILOTs) within the high accretion power model \citep{SokerKashi2016}. Also, they are a key ingredient in the shaping of planetary nebulae by the grazing envelope evolution \cite{Soker2020}.
It was also demonstrated that interaction of jets with a spherical flow can create bipolar planetary nebulae, and can onset instabilities that account for unique features such as the crowns in the Ant Nebula \citep{AkashiSoker2018}.

Jets from an accreting neutron star (NS) companion to a giant star can cause the giant go SN, according to the a common envelope jets supernova (CEJSN) model \citep{Gilkisetal2019}. This is a process very similar to the one that created the GE of $\eta$ Car, except that the companion in \etc is a massive star, rather than a NS.
As also noted by \cite{Gilkisetal2019}, the interaction of a NS with a giant envelope might be an evolutionary stage of the progenitors of most NS-NS binary systems, that will later merge.

It was also suggested that during the explosion of core-collapse SNe, jets interacting with the stellar atmosphere can inflate polar bubbles that will expand faster than the equator, and will result in a rapid collapse when reaching a break point \citep{KaplanSoker2020a} 
This interaction affects the time-dependent optical depth, which may manifest in the resulted light-curve. Jets may also explain late peaks in core-collapse SNe light-curves \citep{KaplanSoker2020b}. 

The effect of jets interaction with gas in supernova impostors was previously studied by \cite{TsebrenkoSoker2013}.
They numerically studied the propagation of the jets through
the previously ejected shell for the supernova impostor SN~2009ip, and found that it reaches velocities of up to $16\,000 \kms$.
The direction of the fast flow in the simulations of \cite{TsebrenkoSoker2013} was similar to the original direction of the jet, while in our simulations here we obtain a perpendicular fast flow.

One effect neglected here is the opening that the secondary wind does in the primary wind.
This is however negligible as during present day periastron passages the secondary wind is suppressed and accretion occurs, let alone this happened during the GE.
In other words the secondary wind was insignificant for that purpose during the GE, and the spherically symmetric density profile adopted is a very good approximation.

Our result show that the high velocity gas in the GE does not require the existence of a triple star system. The accretion model that involves only the known two stars, and does not assume or require a third component, is sufficient to account for the observations of high velocity gas in light echos.
It is possible that the \etc system had a third component and gravitational interaction made the system eccentric, but after it had done so its role is over, and its presence is not essential for the eruption.
The implication of not needing a triple system is that giant eruptions such as the GE are probably more common among evolved very massive stars.

Much more can be learned with similar high resolution simulations that would cover the $\approx 20 \yr$ duration of the GE, including more effects that may influence the accretion rate onto the secondary and consequently the jets.
Such simulations would give us more information on the structure of the Homunculus and the velocities of the gas.

\section*{Acknowledgments}
We thank Nathan Smith for clarifications regarding the high-velocity gas observations.
We appreciate very helpful comments from Noam Soker and from an anonymous referee.
Numerical simulations presented in this work were performed on the Hive computer cluster at the University of Haifa, which is partly funded by ISF grant 2155/15.
AK acknowledges support from the R\&D Authority, and the chairman of the Department of Physics in Ariel University.

\label{lastpage}
\end{document}